\documentclass[preprint,12pt]{elsarticle}

\usepackage{amsmath,amssymb,graphicx,xcolor,booktabs}
\usepackage{hyperref}
\hypersetup{colorlinks=true,linkcolor=blue,citecolor=blue,urlcolor=blue}

\journal{---}

\begin{document}

\begin{frontmatter}


\title{Fungal Systems for Security and Resilience}

\author[1]{Andrew Adamatzky\corref{cor1}}
\ead{andrew.adamatzky@uwe.ac.uk}
\cortext[cor1]{Corresponding author}
\address[1]{Unconventional Computing Lab, UWE Bristol, United Kingdom}

\begin{abstract}
Modern security, infrastructure, and safety-critical systems increasingly operate in environments characterised by disruption, uncertainty, physical damage, and degraded communications. Conventional digital technologies --- centralised sensors, software-defined control, and energy-intensive monitoring --- often struggle under such conditions. We propose fungi, and in particular living mycelial networks, as a novel class of biohybride systems for security, resilience, and protection in extreme environments. We discuss how fungi can function as distributed sensing substrates, self-healing materials, and low-observability anomaly-detection layers. We map fungal properties --- such as decentralised control, embodied memory, and autonomous repair --- to applications in infrastructure protection, environmental monitoring, tamper evidence, and long-duration resilience. 
\end{abstract}


\end{frontmatter}

\section{Introduction}

Modern security, resilience, and infrastructure-protection systems operate in environments that are increasingly disrupted, uncertain, and physically hostile~\cite{little2003toward,biringer2013critical,qudus2025resilient,radvanovsky2023critical}. Critical infrastructure,
transport networks, environmental monitoring installations, and safety-critical facilities are expected to function under conditions of partial failure, limited maintenance, degraded
communications, and sustained stress~\cite{alcaraz2015critical,johnsen2013risk,berenguer2011advances}. In response, there is growing interest in resilience-oriented design paradigms that prioritise persistence, graceful degradation, and recovery over optimisation and performance alone
\cite{hollnagel2016becoming,prokhorenko2020architectural}. Biological systems offer rich sources of such principles, having evolved to survive continuous disturbance over geological timescales`\cite{kiessling2009geologic,walker2012biology,cracraft1985biological}. Among these systems, fungi --- and in particular filamentous fungi forming mycelial networks --- stand out as uniquely resilient, adaptive, and persistent distributed systems~\cite{andrews1992fungal,brown2023fungal,naranjo2019fungal,coleine2022fungi}.

Fungal mycelium consists of networks of microscopic hyphae that grow, branch, fuse, and retract, forming dynamic, decentralised architectures that lack a central control node~\cite{fricker2007network,fricker2008mycelial,heaton2012analysis,fricker2017mycelium}. These networks
continuously reorganise in response to environmental conditions, damage, competition, and resource availability \cite{fricker2008mycelial,tomavsevic2022ergodic}. Importantly, sensing, processing, memory, and repair are not separated into distinct components but are integrated within the same living substrate. This integration contrasts sharply with engineered systems, where sensing, computation, and actuation are typically modular and vulnerable to interface failure.

Recent advances have revealed that fungi are not merely passive structural organisms but active information-processing systems~\cite{adamatzky2023fungal}. Electrical potential fluctuations and spike-like events have been observed in a wide range of
fungal species and  substrates ~\cite{slayman1976action,olsson1995action,adamatzky2018spiking,adamatzky2022language,dehshibi2021electrical,fukasawa2024electrical,buffi2025electrical,fukasawa2025electrical}.  These signals respond to mechanical disturbance, chemical exposure, nutrient availability, temperature, and injury, indicating that mycelium functions as a continuous, analogue sensing medium~\cite{gow1984transhyphal,harold1985fungi,gow1989relationship,gow1995electric,feng2019analysis,phillips2023electrical,fukasawa2023electrical}. Such properties have motivated research into fungal computation, where living mycelium is examined as a substrate for pattern recognition, logic implementation, and unconventional computing \cite{adamatzky2023fungal}.

In parallel, fungi have emerged as promising functional materials~\cite{manan2021synthesis,aiduang2022amazing,gandia2021flexible,cerimi2019fungi,muiruri2023sustainable}. Mycelium-based composites can be grown into predefined shapes, exhibit favourable strength-to-weight ratios, and possess intrinsic self-healing and damping properties. These materials are biodegradable, low-energy to produce, and capable of responding to damage through changes in mechanical and electrical properties. Unlike inert materials, mycelium-based structures could remain metabolically active for extended periods, enabling sensing and adaptation to be embedded directly within structural components~\cite{adamatzky2023fungal}.

Despite these advances, research on fungi has largely progressed along two separate trajectories. One line of work treats fungi as unconventional computing or sensing substrates, focusing on electrophysiology, information processing, and biohybrid interfaces~\cite{adamatzky2023fungal}. Another line treats fungi as novel sustainable materials for construction, packaging, and insulation~\cite{gandia2021flexible, adamatzky2023fungal, cerimi2019fungi}. 

This work shows that for security and resilience applications, these perspectives should be unified. The most powerful potential of fungal systems lies precisely in their dual nature: fungi are simultaneously
\emph{materials and machines}, combining structural function with sensing, processing, and repairwithin a single living system. This integration is particularly relevant for high-risk and extreme environments, where maintenance access is limited and failure can propagate rapidly. Living fungal systems offer several properties that align closely with resilience requirements. First, they are decentralised and robust to partial destruction: cutting or damaging a region of mycelium does not terminate system function, as flows are rerouted and connectivity is regenerated.
Second, fungal systems operate with minimal energy input and without active emissions, making them well suited to low-observability and long-duration deployments. Third, adaptation occurs through
growth and morphological change instead of software updates, reducing vulnerability to digital attack and configuration drift.

Beyond sensing and material properties, fungi exhibit behaviours that can be described as proto-cognitive~\cite{money2021hyphal, adamatzky2023fungal, ostos2025mind}. Experimental studies have demonstrated habituation to repeated stimuli, sensitisation to novel or harmful events, and temporal pattern sensitivity in fungal electrical responses. These behaviours suggest that fungi
can maintain internal baselines and adjust responses based on experience, without requiring explicit symbolic representations or training phases. For resilience-oriented systems, such implicit learning is advantageous, as it enables long-term adaptation under conditions where data, labels, and retraining are unavailable.

The contributions of our paper are threefold. First, we synthesise biological, electrophysiological, and materials-science perspectives on fungi into a unified framework for security and resilience
applications. Second, we articulate how living fungal networks can function simultaneously as sensing substrates, proto-computational systems, and self-healing materials. Third, we outline a research roadmap for developing biohybrid fungal systems that complement conventional digital technologies, offering orthogonal layers of protection and resilience instead of replacement.

Bio-inspired security and resilience research has traditionally focused on neural systems (learning and classification), immune systems (detection and response), and swarm systems (coordination)\cite{johnson2020bio,mthunzi2019bio,myakala2025artificial,bitam2016bio,ahsan2020applications}. While valuable, these paradigms are often software-defined and assume stable computational resources, connectivity, and energy supply.  This work does not propose fungi as replacements for digital security or sensing systems, but as complementary substrates that integrate sensing, processing, memory, and repair within a persistent physical network and fail differently.

\section{Fungi as distributed resilient systems}

\subsection{Mycelial architecture}

The defining structural feature of filamentous fungi is the mycelial network: an interconnected system of microscopic hyphae that grows by apical extension, branching, fusion (anastomosis), and selective regression~\cite{fricker2017mycelium}. 

\begin{figure}[!tbp]
    \centering
    \includegraphics[width=0.5\linewidth]{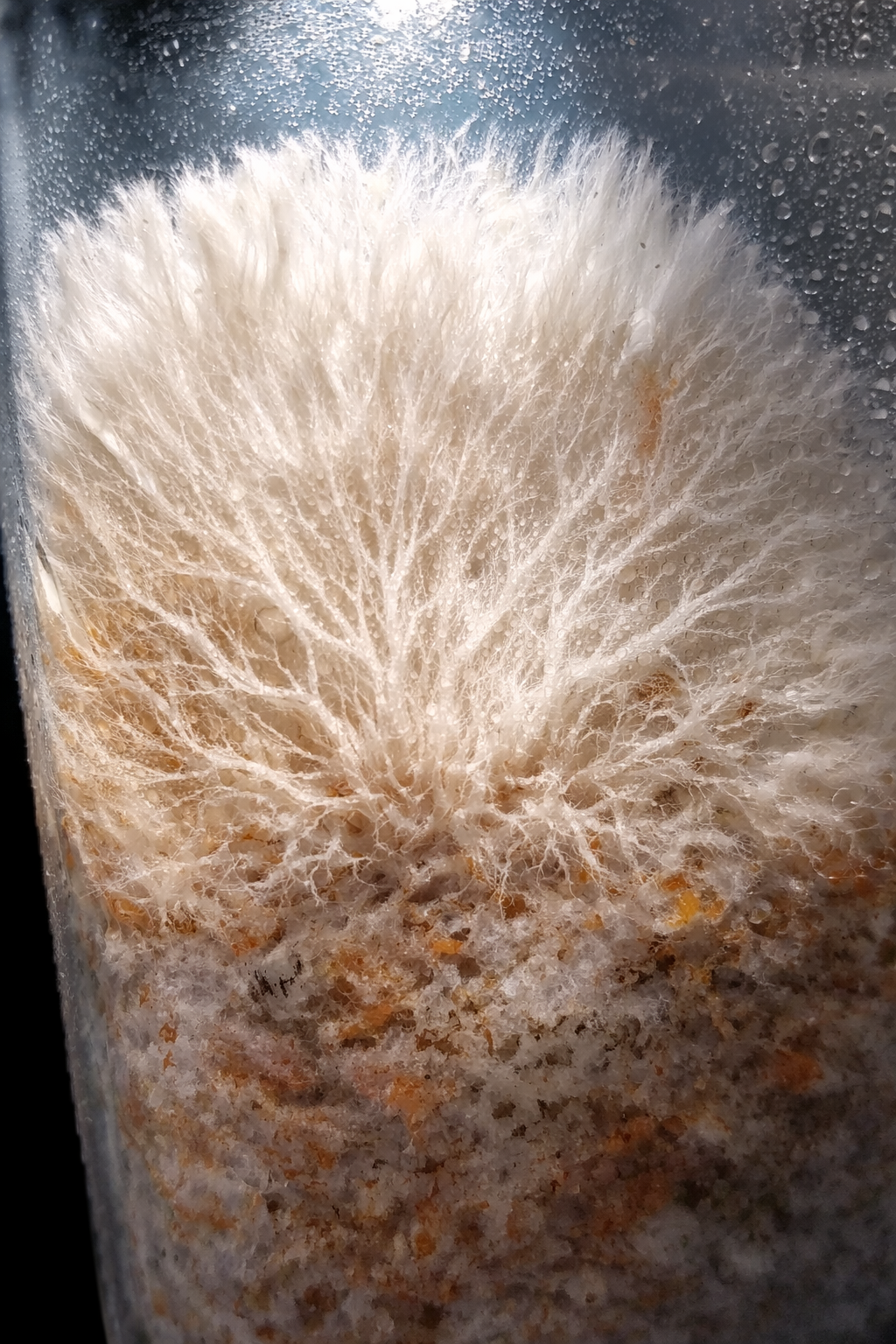}
    \caption{Enhanced photo of a mycelium network shows wave like propagation from the substrate along the wall of a glass container. }
    \label{fig:mycelium}
\end{figure}

Unlike engineered networks that are typically designed, deployed, and then maintained against
change, mycelial architecture is inherently dynamic~\cite{fricker2008mycelial} (Fig.~\ref{fig:mycelium}). Network topology is continuously remodelled in response to environmental conditions, resource availability, mechanical stress, and damage~\cite{branco2022mechanisms}. There is no central coordinating element; instead, global organisation emerges from local growth rules and physiological feedback.

Decentralised control is a key contributor to fungal resilience. Each hyphal tip acts as a semi-autonomous growth and sensing unit, responding locally to chemical, physical, and nutritional cue~\cite{goriely2008mathematical,lopez1994pulsed}, similar to plant root apex~\cite{baluvska2021root}. Decisions about where to grow, branch, thicken, or retract are made without reference to a global plan~\cite{harris2008branching}. As a result, the network has no single point of failure. Damage to one region does not propagate as system-wide collapse but is absorbed locally, with neighbouring regions adapting their growth and transport patterns to compensate.

Redundancy is intrinsic to mycelial networks~\cite{fricker2008interplay}. Hyphae frequently form  multiple parallel connections, providing alternative pathways for transport and signalling. When a pathway is severed or obstructed, flows of cytoplasm, nutrients, and electrical activity can be rerouted through existing connections or newly formed branches. This contrasts with many engineered monitoring systems, where
redundancy must be explicitly designed and often remains limited by cost, weight, or complexity constraints.

Mycelial topology is not static but adaptive~\cite{fricker2008mycelial,fricker2017mycelium, du2018lattice,islam2017morphology}. Regions that consistently support resource transport tend to thicken and become more conductive, while underutilised connections regress and may eventually be abandoned~\cite{park1999effect,nilsson2005growth}. This process of continuous growth and pruning allows the network to balance efficiency with robustness. Instead of optimising for a single configuration, the mycelium maintains a flexible structure capable of responding to changing conditions. Such behaviour mirrors principles advocated in resilience engineering, where adaptability and reconfigurability are prioritised over fixed optimisation.

Importantly, sensing and structure are inseparable in mycelial architecture~\cite{adamatzky2021adaptive}. The  same physical network that provides mechanical cohesion and transport also serves as the sensing substrate. Mechanical deformation, compression, or damage directly alters network geometry and local physiological state, which in turn affects transport dynamics and electrical activity~\cite{adamatzky2022living}. As a result, structural change is
itself a form of signal, as e.g. in space exploration by fungi~\cite{held2019intracellular,hanson2006fungi,held2008examining}. This tight coupling between form and function enables mycelial systems to detect and respond to stress without dedicated sensing elements or interfaces.

From the perspective of security and resilience, the mycelial architecture offers a model for distributed monitoring systems embedded directly within materials and environments. A fungal network integrated into soil, walls, or structural components does not require precise placement or calibration of individual nodes.
Coverage emerges organically as the mycelium grows and explores available space. Damage to the structure does not necessarily disable sensing; instead, it becomes part of the network’s input, influencing subsequent growth and signalling.

Finally, mycelial growth operates over long timescales, which is often viewed as a limitation relative to digital systems~\cite{parker1955fairy,zotti2025fungal}. In resilience-oriented applications, however, this slowness can be advantageous. Gradual growth and pruning allow the network to integrate environmental conditions over extended periods, smoothing
short-term fluctuations while remaining sensitive to sustained change. This temporal integration aligns with the requirements of long-duration monitoring, where the objective is not rapid reaction but persistent awareness and graceful adaptation.

Mycelial architecture embodies a set of structural principles—decentralisation, redundancy, adaptability, and tight coupling between structure and sensing—that closely align with the goals of resilient system design. Rather than serving as a biological curiosity, the mycelial network can be understood as a naturally evolved  template for persistent, damage-tolerant, and adaptive monitoring substrates in hostile or uncertain environments.

\subsection{Electrophysiology and signalling}

\begin{figure}[!tbp]
    \centering
    \includegraphics[width=1\linewidth]{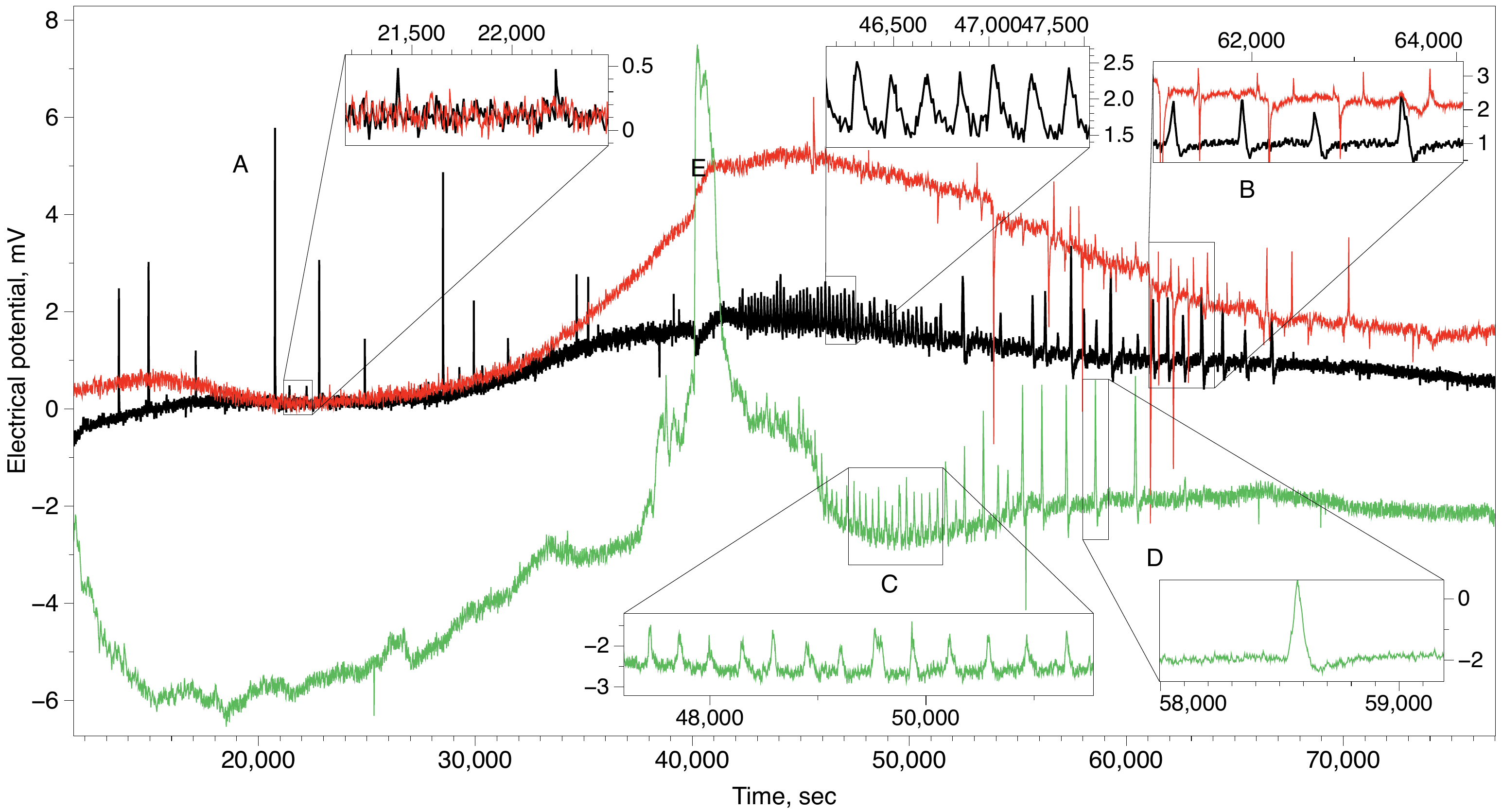}
    \caption{Examples of spiking actvity recorded from a substrate colonised with oyaster fungi.}
    \label{fig:spikes}
\end{figure}

A defining and increasingly well-characterised property of fungal systems is their capacity to generate, propagate, and modulate electrical signals across mycelial networks~\cite{adamatzky2018spiking,buffi2025electrical,phillips2024electrical,geara2023mycelium}. Far from being electrically inert, fungi exhibit rich electrophysiological dynamics that include discrete spike-like events, oscillatory activity, and travelling voltage waves~\cite{adamatzky2018spiking,dehshibi2021electrical,adamatzky2021electrical} (Fig.~\ref{fig:spikes}). These phenomena have been documented across multiple species and experimental settings, and are now recognised as a core component of fungal sensing, coordination, and adaptation
mechanisms~\cite{adamatzky2023fungal}.

Electrical activity in fungi typically manifests as transient changes in
extracellular or intracellular potential, with amplitudes on the order of
millivolts and durations ranging from seconds to minutes~\cite{slayman1965electrical,gow1984transhyphal,gow1995electric}. While fungal spikes
are slower and broader than neuronal action potentials, they are not random
fluctuations. Instead, they form structured temporal patterns that correlate
with physiological processes such as nutrient transport, growth, injury
response, and environmental sensing. Spike trains often exhibit repeatable signatures
associated with specific stimuli, including mechanical perturbation, chemical
exposure, changes in humidity, temperature shifts, and light intensity.

\begin{figure}[!tbp]
    \centering
    \includegraphics[width=0.8\linewidth]{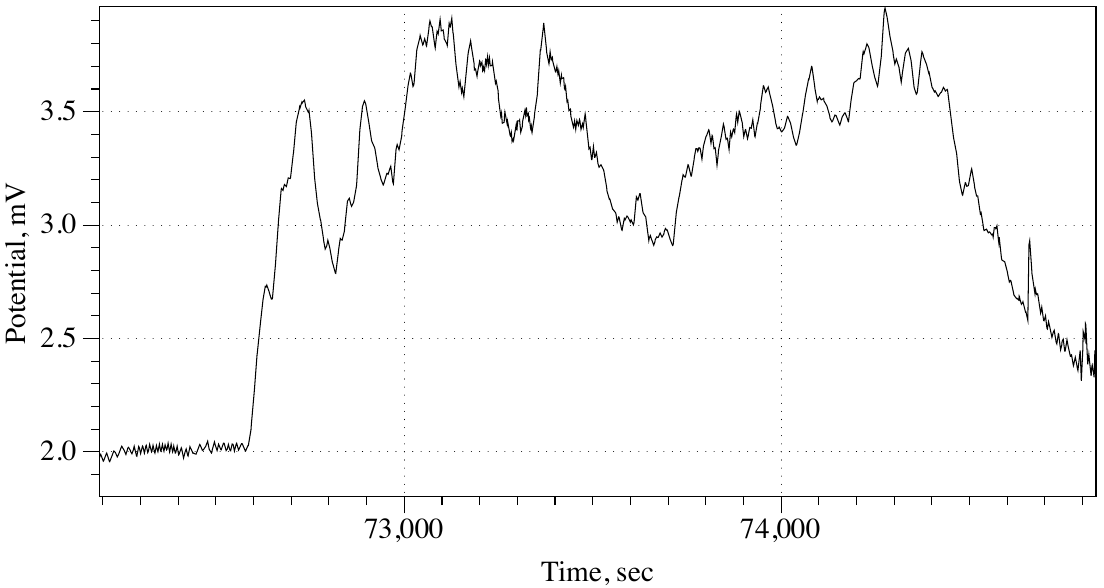}
    \caption{Electrical response of fungal composite to stretching.}
    \label{fig:stretching}
\end{figure}

Mechanical disturbance is one of the most robust triggers of fungal electrical
responses. Compression, bending, cutting, or vibration of hyphae induces bursts
of electrical activity that propagate away from the site of disturbance through
the mycelial network~\cite{adamatzky2022living} (Fig.~\ref{fig:stretching}). Importantly, the spatial extent and temporal structure of
these responses depend on both the magnitude of the disturbance and the prior
physiological state of the network. Minor, repeated perturbations often produce
diminishing responses, while novel or damaging events elicit larger and more
persistent electrical signals. This behaviour supports the interpretation of
fungal electrophysiology as an adaptive sensing mechanism.

Chemical and metabolic stimuli also modulate fungal electrical dynamics.
Exposure to nutrients, toxins, or signalling compounds produces characteristic
changes in spiking frequency and waveform~\cite{dehshibi2021stimulating,adamatzky2021reactive}. In laboratory studies, gradients of
nutrients have been shown to bias the directionality of propagating electrical
waves, effectively coupling electrophysiology to resource acquisition and
growth decisions~\cite{adamatzky2021reactive}. Such coupling suggests that electrical activity plays a role
in coordinating distributed metabolic processes across spatially extended
networks, allowing fungi to integrate local chemical information into coherent
global responses.

Oscillatory electrical behaviour is another prominent feature of fungal
systems~\cite{slayman1965electrical,gow1995electric,adamatzky2018spiking}. Slow oscillations, often linked to rhythmic transport of cytoplasm and
ions, can persist over long durations and synchronise activity across distant
regions of the mycelium~\cite{adamatzky2023multiscalar}. These oscillations interact with faster spike events,
creating multi-timescale electrical dynamics. From an information-processing
perspective, this hierarchy enables fungi to encode both rapid perturbations
and long-term environmental trends within the same signalling substrate.

Crucially, electrical signalling in fungi is inherently distributed. Signals do
not originate from a specialised organ or central node but emerge from the
collective activity of many interacting hyphae. Propagation occurs through the
continuous cytoplasmic and membrane-connected network, with branching and fusion
points shaping signal attenuation, amplification, and delay. As a result,
signal transmission is sensitive to network topology and physiological state,
linking electrophysiology directly to mycelial architecture.

Fungal electrical activity exhibits properties analogous to computation in
excitable media~\cite{adamatzky2023multiscalar,keener1980waves,ostojic2011spiking,scheibner2024spiking}. Spike trains can interact, interfere, and annihilate; waves can
merge or split at junctions; and network morphology constrains signal routing.
These behaviours have been examined as substrates for unconventional computing,
including logical operations and pattern recognition~\cite{adamatzky2020boolean,beasley2021electrical}. For the purposes of
security and resilience, however, the significance of these findings lies less
in computation per se and more in their implications for adaptive sensing and
context awareness.

Unlike digital sensors that produce discrete, pre-defined outputs, fungal electrophysiology provides a continuous analogue representation of system state. Electrical signals reflect the integrated influence of mechanical stress,  chemical conditions, damage, and metabolic activity over time. This makes mycelium particularly well suited to detecting anomalies and transitions rather
than precise measurements. Gradual changes in baseline electrical behaviour can indicate emerging structural fatigue, environmental degradation, or sustained disturbance long before catastrophic failure occurs.

Another important characteristic is the low observability of fungal electrical signalling. Signals are internal to the biological substrate and do not require active emission into the environment. Any external interface—such as electrodes used for monitoring—can be minimal, passive, and intermittent. This stands in contrast to many electronic sensing systems that rely on continuous transmission
and are therefore susceptible to detection, interference, or spoofing.

From a resilience engineering perspective, fungal electrophysiology supports graceful degradation. Damage that disrupts part of the network alters signal paths and patterns but rarely eliminates electrical activity entirely. Instead, the system continues to generate and propagate signals through remaining
connections, providing degraded but still meaningful information about system state. Over time, regrowth and reconfiguration can partially restore signalling capacity without external intervention.

Fungal electrophysiology constitutes a biologically evolved,
distributed signalling system that integrates sensing, memory, and adaptation within a single living substrate. Electrical spikes, oscillations, and propagating waves enable mycelial networks to respond to environmental change, coordinate activity across space, and maintain awareness under damage and uncertainty. For security and resilience applications, these properties position fungal mycelium as a continuous analogue sensing medium capable of long-duration,
low-observability monitoring in environments where conventional electronic systems struggle to persist~\cite{thompson1999analysis,cressler2017extreme,hassan2018electronics}.

\subsection{Self-healing and persistence}

Self-healing is a defining feature of fungal systems and a primary reason for
their exceptional persistence under adverse conditions~\cite{elsacker2023fungal,li2025mycelium,wang2024fabrication,elsacker2024biological}. Unlike engineered
structures that rely on external maintenance or replacement of failed
components, mycelial networks possess intrinsic mechanisms for damage detection,
containment, and repair. These processes operate continuously and autonomously,
allowing fungal systems to survive partial destruction and prolonged stress
without central coordination.

When hyphae are mechanically damaged—through cutting, compression, abrasion, or fracture—fungi rapidly isolate the affected region. Cytoplasmic flow is sealed locally through septal closure and membrane reconfiguration, preventing loss of cellular contents and limiting the spread of  damage~\cite{markham1987woronin,van2009cytoplasmic,roper2019mycofluidics}. This immediate response stabilises the surrounding network and preserves functionality in adjacent regions. Electrical activity and transport processes are altered near the injury site, effectively marking the damaged area within the physiological state of the
network.

Beyond short-term containment, fungi initiate longer-term repair through growth and reconfiguration. New hyphal branches emerge from intact regions, often bypassing damaged zones and reconnecting previously linked areas. Because mycelial networks contain loops and redundant pathways, transport and signalling can be rerouted even before physical regrowth occurs. Over time, regenerated
connections may restore partial or full network integrity, although often with altered topology that reflects the history of damage.

Crucially, self-healing in fungi does not aim to restore an original configuration. Instead, repair is adaptive and context-dependent. Regrown regions may differ in hyphal thickness, conductivity, or branching density, embedding information about past stress within the structure of the network itself. This stands in contrast
to engineered self-healing materials that seek to erase damage and return to a prior state. In fungal systems, damage leaves a persistent imprint that shapes future behaviour, contributing to embodied memory and adaptive sensitivity.

Persistence in fungal systems extends beyond mechanical repair. Many fungi can withstand prolonged periods of desiccation, nutrient scarcity, temperature extremes, and chemical exposure by entering metabolically reduced  states~\cite{magan2007fungi,zak2004fungi,coleine2022fungi}.
Activity can resume when conditions improve, often without external intervention. This capacity for dormancy and reactivation enables fungal networks to function across discontinuous operational timelines, a property rarely achievable in conventional sensing or monitoring technologies.

From a resilience engineering perspective, these properties support graceful 
degradation not the catastrophic failure. As damage accumulates, system
performance may decline gradually—through reduced signal clarity, slower
responses, or partial loss of coverage—but meaningful function is retained~\cite{phillips2023electrical}.
Importantly, failure does not propagate uncontrollably through the network.
Instead, localised damage is absorbed and compensated for through reconfiguration
and regrowth.

In biohybrid materials, mycelium-based composites inherit aspects of this
self-healing behaviour~\cite{van2021review,menon2019screening,luo2018interactions,zhang2021study,elsacker2023fungal}. While structural repair is limited by environmental
conditions and substrate availability, living mycelial components can respond to
internal cracking or deformation through changes in electrical and mechanical
properties. Such responses provide intrinsic damage sensing while extending the
functional lifetime of the material. Even when full structural recovery is not
possible, the system retains awareness of its degraded state.

For security and resilience applications, self-healing and persistence reduce
dependence on continuous maintenance, external repair, and precise environmental
control~\cite{psaier2011survey,ekeocha2021challenges,zhou2022self}. A fungal sensing or protective layer embedded within infrastructure does
not require immediate intervention following damage to remain informative.
Instead, the system continues to operate in a degraded but meaningful mode,
encoding both current conditions and damage history within its structure and
physiology.

Fungal self-healing is not merely a material property but a systemic
capability arising from decentralised architecture, adaptive growth, and
embodied memory. These mechanisms enable fungal systems to survive damage,
reconfigure function, and persist under sustained disruption. Such behaviour
aligns closely with the goals of resilient system design, where endurance and
recoverability are prioritised over uninterrupted performance.

\subsection{Biological variability and operational robustness}

Any practical deployment of living fungal systems must explicitly account for biological variability~\cite{he2022species}. Fungal behaviour is inherently species-dependent and sensitive to growth conditions, including substrate composition, temperature, moisture, nutrient availability, and age of the mycelial network. Electrical activity, growth dynamics, and adaptive responses can therefore vary across species and environmental contexts, and may drift over time as physiological state changes. From a conventional engineering perspective, such variability is often treated as a limitation or source of uncertainty.

In resilience-oriented applications, however, biological variability is not necessarily detrimental and may in fact be advantageous. Fungal systems do not rely on fixed thresholds or calibrated outputs; instead, they establish baselines through continuous interaction with their environment. Variability in growth rate, conductivity, or electrical signalling becomes part of the system’s internal reference, allowing it to differentiate between routine fluctuations and novel or sustained disturbances. This baseline learning is particularly valuable in long-duration deployments where environmental conditions change seasonally or unpredictably.

Coupling between fungal physiology and environmental factors such as temperature and moisture further enhances sensitivity to meaningful change \cite{braunsdorf2016fungal, alonso2009fungi}. Not attempting to isolate or compensate for such coupling, fungal systems integrate it, enabling detection of compound or slowly evolving conditions that may elude conventional sensors. Similarly, species-specific differences in growth and signalling offer opportunities for tailoring fungal substrates to particular operational environments, instead of enforcing uniform behaviour across deployments.

For security and tamper-evident applications, biological variability increases resistance to spoofing and reset. Because system responses emerge from a living, history-dependent substrate, reproducing or erasing a specific physiological state is impractical. Disturbances are therefore encoded relative to each system’s unique developmental trajectory, strengthening the persistence and authenticity of embodied records.

From this perspective, biological variability should be understood not as noise to be eliminated, but as a source of adaptive richness that supports baseline formation, anomaly detection, and long-term resilience under uncertain and degraded conditions.

\section{Biohybrid fungal systems for security and protection}

\subsection{Living fungal sensors}

Living fungal networks provide a sensing paradigm fundamentally different from
conventional electronic sensors. Rather than measuring isolated physical
quantities through dedicated transducers, mycelium functions as a continuous,
distributed sensing medium whose electrical, physiological, and structural
states are directly shaped by its environment. Mechanical stress, chemical
exposure, light, and electrical perturbations all induce characteristic and
reproducible changes in fungal activity, allowing mycelium to act as an
integrated, multi-modal sensor when embedded within materials or landscapes.

\paragraph{Mechanical sensing}
Mechanical sensitivity is one of the most robust and well-documented properties
of fungal systems. Experimental studies have shown that compression, stretching,
cutting, vibration, and impact applied to hyphae or mycelial mats trigger
distinct electrical responses, including bursts of spike activity, changes in
oscillation frequency, and shifts in baseline potential~\cite{adamatzky2022living,adamatzky2021towards}. These responses are not
confined to the point of contact: electrical signals propagate through the
network, enabling detection of mechanical events at locations distant from the
disturbance~\cite{adamatzky2021reactive}.

Importantly, fungal mechanical sensing integrates both magnitude and persistence.
Transient, low-amplitude vibrations tend to produce brief, localised electrical
responses, while sustained or repetitive stress leads to prolonged signalling
and structural adaptation. This makes mycelium particularly well suited to
monitoring slow deformation, fatigue, or progressive damage in structures such
as walls, tunnels, or embankments, where early changes are spatially distributed
and difficult to localise with point sensors.

\paragraph{Chemical sensing}
Fungal electrical activity is highly sensitive to chemical conditions in the
surrounding environment. Changes in nutrient availability, pH, moisture, and the
presence of toxic or unfamiliar compounds all modulate spiking behaviour and
oscillatory dynamics~\cite{adamatzky2021reactive}. Different classes of chemical stimuli produce distinct
temporal signatures, reflected in spike frequency, waveform shape, and recovery
time.

While fungi do not identify chemical species with analytical precision, they could be effective at detecting deviations from established baselines. For
example, gradual exposure to benign nutrients produces responses that differ
qualitatively from those induced by harmful or inhibitory compounds. This
property enables fungal systems to function as anomaly detectors for leakage,
contamination, or chemical intrusion in soil, infrastructure, or enclosed
spaces, without requiring prior knowledge of the specific substance involved.

\paragraph{Optical and photic sensitivity}
Although fungi lack specialised visual organs, many species exhibit measurable
responses to light~\cite{beasley2023fungal,adamatzky2021towards}. Exposure to changes in illumination, wavelength, or light
intensity has been shown to alter electrical activity and
oscillatory patterns in mycelial networks~\cite{adamatzky2021towards}. These responses are often slow and
integrative, reflecting cumulative exposure instead of instantaneous changes.

Such photic sensitivity is relevant in contexts where light exposure correlates
with environmental change or human activity. For instance, unexpected light
penetration into normally dark subsurface environments may signal structural
breach or unauthorised access. When embedded within walls or underground
structures, fungal networks can register these changes indirectly through
altered electrical dynamics, even when light exposure is diffuse or intermittent.

\paragraph{Electrical coupling and perturbation sensing}
Fungal systems are themselves electrically active, and their endogenous
electrophysiology is sensitive to externally applied electric fields and
potentials. Weak electrical stimulation can entrain or suppress spiking
activity, alter oscillation patterns, and modify signal propagation across the
network. Conversely, changes in the electrical environment—such as grounding
faults, electrostatic discharge, or nearby powered equipment—can perturb fungal
electrical behaviour.

This bidirectional sensitivity enables mycelium to function as a passive monitor
of local electrical conditions. In embedded applications, fungal networks can
register changes in electrical fields or conductivity associated with moisture
ingress, material degradation, or tampering with electrical infrastructure.

\paragraph{Distributed and analogue sensing}
Across all modalities, a key feature of fungal sensing is its distributed and
analogue nature. Signals do not correspond to discrete measurements but to
patterns of activity distributed across space and time. Mechanical, chemical,
optical, and electrical inputs interact within the same biological substrate,
producing composite responses that reflect the integrated state of the
environment.

This contrasts to conventional sensor arrays, where different modalities
are handled by separate devices and fused digitally. In fungal systems, sensing
fusion occurs intrinsically, through physiological coupling and network dynamics.
As a result, fungal sensors are particularly effective in detecting transitions,
anomalies, and sustained changes.

\paragraph{Energy efficiency and low observability}
Living fungal sensors operate continuously with minimal external energy input.
Electrical activity arises from endogenous metabolic processes, and sensing does
not require active emission of signals into the environment. External interfaces
can therefore be passive, low-power, and intermittent. This low observability
makes fungal sensing layers difficult to detect, probe, or disrupt remotely,
which is advantageous in security-sensitive environments.

\paragraph{Implications for resilient sensing}
When embedded in soil, walls, or composite materials, mycelium transforms the
structure itself into a sensing entity. Coverage arises organically as the
network grows, and damage to the sensing substrate becomes part of the sensed
signal not a cause of failure. The result is a sensing system that is
persistent, adaptive, and tolerant of partial destruction.

\subsection{Low-observability anomaly detection}

A distinctive advantage of fungal sensing systems is their inherently low
observability. Unlike most conventional monitoring technologies, fungal systems
do not rely on active signal emission, continuous communication, or externally
addressable digital interfaces. As a result, they present a markedly reduced
surface for detection, probing, interference, or spoofing, making them well
suited to tamper-resistant monitoring in contested, sensitive, or degraded
environments~\cite{skorobogatov2006tamper,puesche2018concept}.

Most electronic sensors operate by emitting or exchanging signals—radio
transmissions, optical pulses, acoustic waves, or periodic network traffic—that
can be detected, intercepted, or jammed~\cite{eren2018wireless,jena2018wireless,fraden2004handbook,osanaiye2018statistical,li2007optimal}. Even passive sensors are typically
connected to identifiable cabling or digital infrastructure that reveals their
presence and function. In contrast, fungal sensing is intrinsically internal.
Mechanical, chemical, optical, or electrical perturbations are registered as
changes in the physiological and electrophysiological state of the mycelium
itself, without the need for external interrogation or continuous data
transmission.

This internalisation of sensing significantly complicates remote probing. There
is no standard protocol to interrogate, no addressable node to query, and no
predictable timing or signature that reveals when sensing is occurring. Any
external readout—such as electrodes or conductive contacts—can be minimal,
localised, and operated intermittently, further reducing detectability. In many
deployments, fungal sensing layers may be indistinguishable from inert materials
or naturally occurring biological growth.

Low observability also confers resistance to spoofing. Conventional sensors can
often be deceived by injecting false signals, replaying recorded data, or
manipulating input channels~\cite{illiano2015detecting,choi2022survey,sikder2021survey,yu2015integrated}. Fungal systems, by contrast, respond directly to
physical and chemical interactions with their environment. To spoof a fungal
sensor, an adversary would need to reproduce the actual mechanical stress,
chemical exposure, or structural change that the system is designed to detect.
This requirement raises the cost and complexity of deception, particularly in
distributed or subsurface deployments.

Anomaly detection in fungal systems is therefore grounded in embodied change. Deviations from baseline conditions
manifest as altered electrical dynamics, growth patterns, or conductivity within
the living network. Because these baselines are established biologically and
continuously updated through adaptation, they are difficult to predict or
manipulate externally. Gradual drift in environmental conditions is absorbed into
the baseline, while novel or escalating disturbances produce responses that stand
out against the system’s own history.

This mode of anomaly detection is especially valuable in environments where
conventional sensors are easily disrupted. In underground infrastructure,
electromagnetically noisy industrial sites, or disaster zones, radio-based
systems may be unreliable or unusable. Fungal sensing layers embedded in soil,
walls, or structural materials continue to function without reliance on external
communication channels, providing local awareness even when broader monitoring
networks fail.

Another important aspect of low-observability anomaly detection~\cite{shi2020early,li2015parameter} is persistence.
Because fungal systems do not depend on continuous connectivity or active polling,
they can remain dormant or operate in a reduced state for extended periods.
Anomalous events that occur during communication outages or power loss are not
necessarily missed; instead, they may leave lasting physiological or structural
imprints that can be detected upon later inspection. This contrasts with digital
systems, where missed data is often irretrievable.

From a security perspective, fungal anomaly detection aligns with defensive
monitoring goals that prioritise awareness over intervention~\cite{garcia2007design,ashok2014cyber,nunes2014designing,bejtlich2004tao}. Fungal systems provide
persistent evidence that something has changed. This evidence can be interpreted
locally or extracted intermittently and integrated with higher-level digital
systems when conditions permit.

\subsection{Self-healing protective materials}

Mycelium-based composites represent a distinct class of protective materials in
which structural function, sensing, and adaptation are integrated within a
single living substrate~\cite{vandelook2021current,yang2021material,french2023mycelium,al2023mycelium}. Unlike conventional composites that are inert once
manufactured, mycelium-based materials might retain (when necessary) metabolic activity for extended
periods, enabling them to respond dynamically to stress, damage, and
environmental change. This capacity fundamentally alters how protection and
durability can be conceived in resilience-oriented systems.

A key property of live mycelium-based composites is their ability to detect internal
damage through changes in electrical and physiological state. Mechanical
deformation, cracking, compression, or delamination alters hyphal connectivity,
cytoplasmic flow, and ionic transport within the material. These changes manifest
as shifts in electrical activity, including altered baseline potentials,
disrupted oscillatory patterns, or localised spiking responses. As a result,
damage that may not be visible externally can be detected intrinsically, without
the need for embedded strain gauges or inspection-based monitoring.

Structural adaptation further distinguishes mycelium-based materials from inert
alternatives. When subjected to sustained stress or repeated loading, fungal
networks can locally thicken hyphae, modify branching density, or redirect growth
along stress-bearing pathways. Although such adaptation does not restore
original material properties in an engineered sense, it can redistribute load
and slow the progression of damage. This behaviour parallels biological
strategies for reinforcing frequently stressed regions not maintaining
uniform strength throughout a structure.

Partial functional recovery through regrowth is another important aspect of
self-healing. When cracks or fractures disrupt continuity within a mycelium-based
composite, surrounding hyphae may grow into damaged regions, restoring some
degree of connectivity and mechanical cohesion. While the extent of regrowth
depends on environmental conditions such as moisture, temperature, and nutrient
availability, even limited regeneration can extend the service life of protective
structures and preserve sensing capability.

Importantly, self-healing in mycelium-based materials does not imply complete
restoration or invisibility of damage. Instead, repair processes are typically
asymmetric and history-dependent, leaving persistent structural and electrical
signatures. From a security and resilience perspective, this is advantageous.
Not erasing evidence of damage, the material encodes it, providing both
continued protection and long-term awareness of stress history or interference.

Potential applications of such self-healing protective materials include
temporary shelters, protective enclosures, blast or impact-damping barriers, and
containment structures in environments where maintenance access is limited.
Mycelium-based panels or coatings can be grown into specific shapes and
integrated with conventional materials such as timber, concrete, or polymer
matrices, providing hybrid structures that combine biological adaptability with
engineered strength.

In emergency or disaster-response contexts, self-healing protective materials
offer particular advantages. Structures assembled rapidly from mycelium-based
components can continue to adapt after deployment, responding to settlement,
moisture ingress, or mechanical damage without immediate repair. Even when
subjected to partial destruction, such materials degrade gradually not
failing abruptly, maintaining a degree of protective function while signalling
their compromised state.

From a resilience engineering standpoint~\cite{raz2019multi,cottam2019defining,rosowsky2020defining}, the value of mycelium-based protective
materials lies not in achieving maximum strength or longevity, but in supporting
graceful degradation, intrinsic damage awareness, and partial recovery. These
materials blur the distinction between structure and sensor, transforming
protective elements into active participants in system resilience.

\subsection{Tamper-evident systems}

Tamper evidence is a critical but often underemphasised component of security and
resilience. In many contexts, the primary requirement is not immediate detection
or automated response, but reliable evidence that interference has occurred.
Living fungal systems provide a distinctive approach to tamper evidence by
encoding disturbance directly into their physical and physiological state. This
evidence is persistent, difficult to erase, and cannot be reset through digital
means.

\begin{figure}[!tbp]
    \centering
    \includegraphics[width=0.8
    \linewidth]{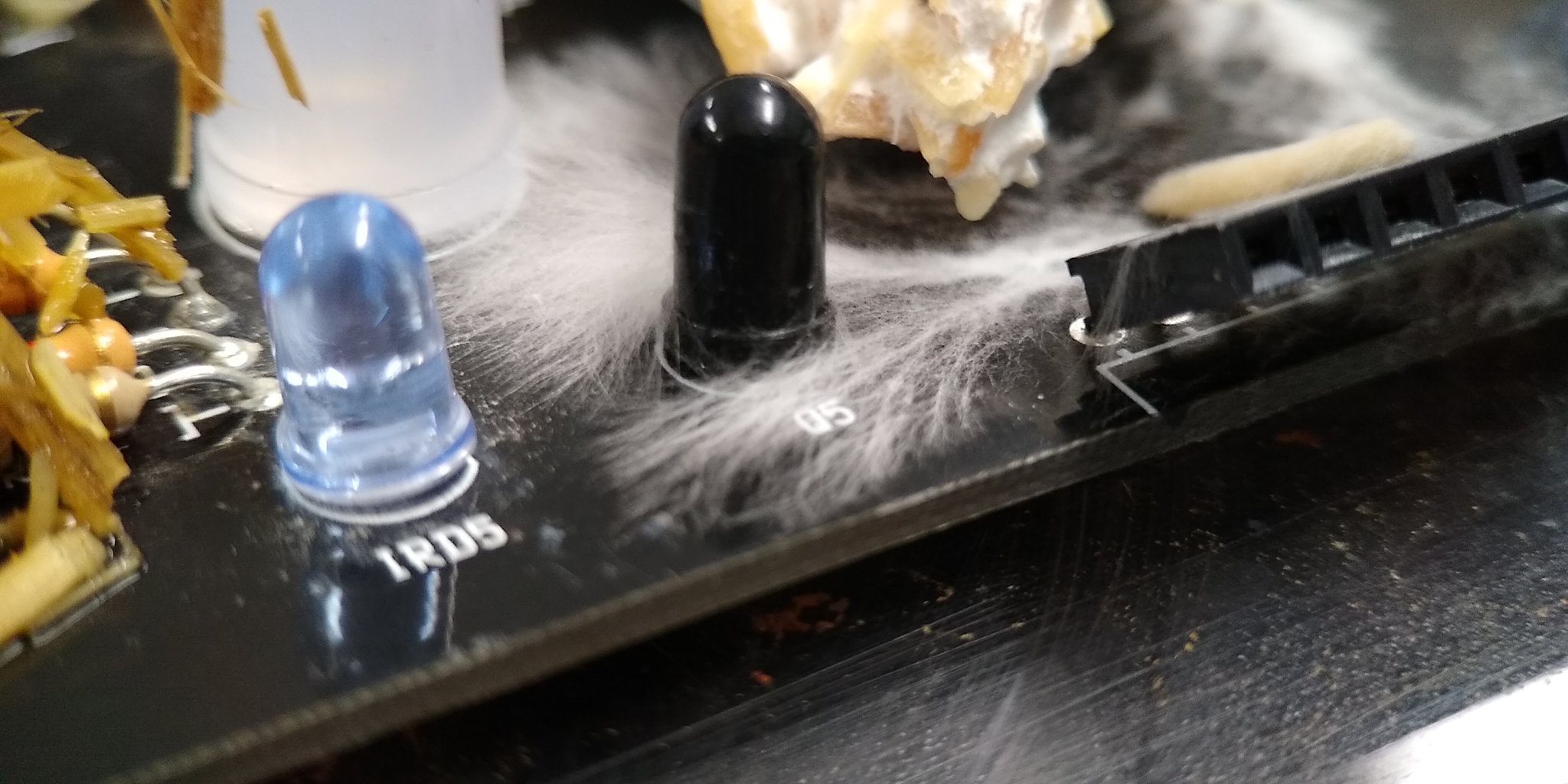}
    \caption{Mycelium propagating on an electronic board.}
    \label{fig:board}
\end{figure}

Living fungal seals and coatings respond to mechanical disturbance, removal,
penetration, or deformation through irreversible changes in network structure,
growth patterns, and electrical behaviour (Fig.~\ref{fig:board}). Cutting, peeling, crushing, or
displacing a mycelial layer alters hyphal connectivity and local physiology in
ways that persist long after the event. Even when the fungal system continues to
grow or partially repair itself, the resulting structure differs measurably from
its undisturbed state, preserving a record of interference.

Unlike electronic tamper sensors~\cite{skorobogatov2011physical,engelbrecht2019design,mosavirik2023impedanceverif,anderson1996tamper,immler2019secure}, which often rely on binary triggers or digital
logs, fungal tamper evidence is analogue and embodied. Disturbance produces
graded changes—such as altered conductivity, modified spiking patterns, or
persistent asymmetries in growth—that reflect both the magnitude and spatial
extent of tampering. This enables post hoc assessment of interference not
simple yes-or-no indication, supporting forensic analysis in situations where
real-time monitoring is unavailable or unreliable.

A key advantage of fungal tamper-evident systems is their resistance to reset or
spoofing. Digital tamper indicators can often be cleared by power cycling,
software manipulation, or component replacement. In contrast, restoring a fungal
seal to its original state would require reconstructing its exact biological
history, including growth trajectory, physiological state, and microstructure.
Such reconstruction is effectively infeasible, particularly in field
deployments. As a result, fungal tamper evidence is robust against both accidental
erasure and deliberate concealment.

Tamper-evident fungal layers are also well suited to distributed and irregular
surfaces. Mycelium naturally conforms to complex geometries, filling gaps,
crevices, and interfaces that are difficult to instrument with conventional
devices. When applied as coatings or embedded within materials, fungal systems
provide continuous coverage. Any attempt to
access protected components necessarily disrupts the living network, leaving
detectable traces.

Potential applications include seals for enclosures containing critical
equipment, coatings on access panels or storage containers, and interfaces
between structural components where unauthorised separation must be detectable.
In such roles, fungal systems function as passive witnesses. They do not prevent access, but they ensure that access cannot occur
without leaving evidence.

\subsection{Evaluation criteria and performance}

A recurring question for any proposed sensing or security technology is how its effectiveness can be evaluated. Fungal systems differ fundamentally from conventional digital sensors in that they do not aim to provide precise, instantaneous measurements, but rather persistent, adaptive awareness under disruption. As such, their evaluation requires criteria aligned with resilience-oriented objectives.

Several qualitative and semi-quantitative evaluation dimensions are particularly relevant for fungal systems deployed in security and resilience contexts. Persistence under damage is a primary criterion: a fungal system can be considered functional if meaningful electrical or physiological activity continues following partial destruction, deformation, or environmental stress. Performance is assessed not by uninterrupted operation, but by the degree to which sensing and signalling degrade gradually.

Signal stability and drift over time constitute a second evaluation dimension. Because fungal systems operate over long timescales, gradual baseline drift is expected and may reflect adaptation. Evaluation therefore focuses on whether such drift remains bounded and interpretable, and whether deviations associated with disturbance or tampering remain distinguishable from long-term physiological change.

A third criterion concerns sensitivity to sustained versus transient perturbations. Fungal systems are particularly well suited to resilience applications when they attenuate responses to brief, routine fluctuations while amplifying responses to persistent or novel disturbances. Experimental evaluation can therefore examine differential electrical responses to repeated low-intensity stimuli versus sustained or escalating stress.

Recovery time following disturbance provides an additional measure of system effectiveness. Recovery need not imply return to an original state; instead, evaluation focuses on the time required for signalling activity, growth, or conductivity to stabilise after damage, and on the degree to which altered structure encodes a persistent record of the disturbance.

Finally, observability and detectability form a critical evaluation dimension in security-sensitive deployments. A fungal system may be considered successful if it remains operational without continuous active emission, network communication, or externally identifiable signatures. Assessment therefore includes not only sensing performance, but also the extent to which the system avoids detection, probing, or spoofing through remote means.

Collectively, these criteria emphasise functional persistence, adaptive differentiation, and low observability over speed or precision. While quantitative benchmarks will be application- and species-specific, articulating these evaluation dimensions provides a practical framework for laboratory characterisation, biohybrid prototyping, and field trials of fungal systems in security and resilience contexts.

\section{Proto-cognitive properties and decision support}

Fungal systems exhibit a range of behaviours that, while not cognitive in the
human or animal sense, are increasingly described as \emph{proto-cognitive} or
\emph{minimal cognitive} processes~\cite{adamatzky2023fungal,adamatzky2022fungi}. These behaviours arise from the interaction
of growth dynamics, electrophysiological signalling, chemical feedback, and
structural adaptation within mycelial networks. Notably, such processes
support decision-relevant functions---prioritisation, filtering, memory, and
context sensitivity---without symbolic representation, centralised control, or
explicit models.

\subsection{Prioritisation and resource allocation}

One of the most robustly documented proto-cognitive properties of fungi is their
ability to prioritise resource flows under constraint. Mycelial networks actively
redistribute nutrients toward regions of higher utility, such as nutrient-rich
zones or actively growing fronts, while withdrawing resources from less productive
areas~\cite{tlalka2008mycelial,watkinson2005new,fricker2008imaging,watkinson2005new,boswell2002functional,boswell2007development}. Fungi can dynamically reweight transport
pathways based on nutrient gradients, damage, or competition, producing network
topologies that approximate solutions to optimisation problems such as shortest
paths and minimum-cost flow.

In a security or resilience context, this behaviour provides a natural analogue to
decision support under scarcity. For example, a fungal sensing substrate embedded
within infrastructure could preferentially change amplitudes of their signals originating from regions
experiencing sustained stress or disturbance, while attenuating background noise
from stable regions. This prioritisation emerges intrinsically from network dynamics,
without requiring predefined thresholds or global optimisation routines.

\subsection{Embodied memory and structural encoding}

Unlike digital systems, fungal memory is not stored symbolically but is embodied in
physical structure and physiological state~\cite{beasley2022mem}. Changes in hyphal thickness, branching
patterns, connectivity, and conductivity persist long after the initiating stimulus
has ceased.

In practical terms, this allows fungal systems to ``remember'' past disturbances or
stress events without maintaining explicit logs or state variables. A mycelium-based
material used as a protective layer may exhibit permanently altered electrical
properties following tampering or damage, providing long-term evidence of interference
even in the absence of continuous monitoring.

\subsection{Decision support without centralised control}

The proto-cognitive behaviours described above collectively enable a form of
distributed decision support. Fungal systems continuously integrate multi-modal
inputs---mechanical, chemical, thermal, and electrical---and respond through
coordinated changes in signalling and structure. No single node determines system
behaviour; rather, responses emerge from local interactions constrained by global
resource availability.

This mode of operation is particularly valuable in environments with intermittent
connectivity, limited power, or restricted maintenance access. Fungal substrates can
operate autonomously for extended periods, providing pre-processed, context-aware
signals to higher-level digital systems only when biologically significant deviations
occur. In this sense, fungi function not as classifiers or controllers, but as
\emph{adaptive front-end filters} that reduce informational burden while preserving
sensitivity to meaningful change.

\subsection{Illustrative scenarios}

Concrete examples of proto-cognitive decision support include:
\begin{itemize}
\item A fungal sensor network embedded in a tunnel wall that gradually amplifies
responses to repeated micro-vibrations associated with subsurface movement, while
ignoring routine traffic noise.
\item A mycelium-based composite panel that permanently alters its electrical
signature following partial fracture, encoding damage history within its material
structure.
\item A distributed fungal substrate that suppresses responses to seasonal humidity
cycles but remains sensitive to anomalous chemical exposure indicative of leakage or
contamination.
\end{itemize}

\section{Application domains}

\subsection{Critical infrastructure protection}

Critical infrastructure such as bridges, tunnels, pipelines, power stations,
and water distribution systems requires continuous monitoring over long
timescales, often in environments that are difficult to access or maintain~\cite{brownjohn2007structural,karbhari2009structural,chang2003health,whittle2013sensor}.
Conventional sensor systems depend on stable power, regular calibration, and
network connectivity, making them vulnerable to degradation, tampering, or
failure.

Living fungal systems offer an alternative approach by acting simultaneously as
structural material, sensing substrate, and adaptive monitoring layer.
Mycelium-based composites can be integrated into concrete, insulation panels,
or protective coatings, where changes in mechanical stress, cracking, moisture,
or chemical exposure induce measurable alterations in electrical activity. Because mycelium naturally spans large areas, a
single continuous network can monitor distributed regions without dense sensor
arrays.

For example, a fungal layer embedded along a tunnel lining could gradually
develop distinct electrical signatures in response to water ingress, material
fatigue, or repeated vibration from traffic. Over time, these signatures form a
baseline against which anomalous patterns—such as structural compromise or
unauthorised access—can be detected. Importantly, the fungal system remains
operational even if portions are damaged, as growth and signalling reroute
around compromised regions.

\subsection{Environmental and geotechnical monitoring}

Environmental and geotechnical monitoring often requires detection of subtle,
slowly evolving processes such as soil subsidence, landslide precursors,
underground excavation, or long-term contamination~\cite{soga2019advances, dunnicliff1993geotechnical,dunnicliff1993geotechnical,gong2019advances}. These processes frequently occur in remote locations where maintaining electronic sensor networks is
logistically challenging.

Fungal mycelium naturally inhabits soil and subsurface environments and is
highly sensitive to mechanical deformation, moisture gradients, and chemical
composition \cite{ivarsson2018fungi,reitner2006fungi,ritz2004interactions,finlay2019fungi}. Embedded fungal networks can therefore function as
living geosensors, responding to minute changes in pressure, strain, or soil
chemistry. Electrical activity propagates through the network when hyphae are
compressed, severed, or exposed to novel compounds.

A practical example is the use of fungal sensing layers in slopes or embankments
to detect early signs of instability. Gradual increases in strain or moisture
can induce systematic changes in fungal signalling long before macroscopic
failure occurs. Unlike point sensors, the fungal network provides spatially
continuous coverage and adapts its sensitivity as environmental conditions
change seasonally or climatically.

\subsection{Emergency and disaster response}

During disasters such as earthquakes, floods, wildfires, or industrial
accidents, digital infrastructure is often among the first systems to fail.
Power loss, communication breakdown, and physical destruction severely limit
situational awareness precisely when it is most needed.

Biohybrid fungal systems can support emergency response by providing resilient,
autonomous monitoring that continues to function in degraded conditions. Because
fungal systems require minimal energy and no continuous connectivity, they can
remain active during and after disruptive events. Electrical responses to
structural collapse, flooding, chemical release, or heat exposure provide
real-time indicators of evolving conditions.

For instance, fungal composites integrated into temporary shelters or emergency
barriers could provide continuous feedback on structural integrity, moisture
penetration, or exposure to hazardous substances. After the acute phase of a
disaster, these same systems can support recovery by monitoring gradual
stabilisation or degradation without requiring immediate replacement or repair.

\section{Comparison with conventional technologies}

The potential value of fungal systems for security and resilience is best understood
not as a replacement for conventional digital technologies, but as a fundamentally
different technological layer that operates according to distinct assumptions about
time, failure, observability, and control. This section compares fungal systems with
established sensing, monitoring, and AI-based approaches across several operational
dimensions, using concrete examples drawn from infrastructure protection, environmental
monitoring, and high-risk environments.

\subsection{Failure modes and survivability}

Conventional digital sensing systems are typically designed for accuracy and efficiency
under nominal conditions. Their failure modes, however, are often abrupt. Loss of power,
network connectivity, clock synchronisation, or a single critical component can render
entire sensor arrays inoperative. Even when redundancy is built in, common-mode failures
such as electromagnetic interference, flooding, fire, or deliberate sabotage can disable
large portions of the system simultaneously.

By contrast, fungal systems exhibit failure modes characterised by gradual degradation
instead of sudden collapsing. A mycelial network has no central processing unit, no single point
of coordination, and no requirement for global synchrony. When a region of the network is
damaged—through cutting, crushing, dehydration, or chemical exposure—local activity may
cease, but neighbouring regions typically reroute growth, electrical signalling, and
resource flow around the damaged area. Over time, connectivity may be partially or fully
restored through regrowth.

A concrete example can be drawn from tunnel monitoring. A conventional system may rely on
distributed accelerometers, strain gauges, and humidity sensors connected via wired or
wireless networks to a central monitoring station. If a section of cabling is severed or
a communication node fails, downstream sensors may be lost entirely. A fungal sensing
layer embedded within the tunnel lining, by contrast, would continue to function locally
even if portions are destroyed. While spatial resolution and signal clarity may degrade,
the system retains partial awareness of structural stress and disturbance, aligning with
resilience goals not performance optimisation~\cite{itani2023local}.

\subsection{Observability, emissions, and tamper resistance}

Most digital monitoring systems are externally observable. Wireless sensors emit radio
signals, wired systems require identifiable cabling, and even passive devices often have
predictable electromagnetic signatures. These properties make them susceptible to
detection, interference, spoofing, and targeted disabling. In adversarial or sensitive
environments, the mere presence of such systems can alter behaviour or attract attention.

Fungal systems operate with near-zero active emissions. Electrical activity remains
confined within the biological substrate and any local readout interface. There is no
inherent radio-frequency transmission, no periodic beaconing, and no standardised protocol
that can be interrogated remotely. As a result, fungal sensing layers are difficult to
locate, probe, or spoof without direct physical access.

Consider tamper monitoring for secure enclosures or access points. A conventional approach
might employ switches, seals, or electronic tamper sensors that trigger alarms when opened.
Such devices can often be bypassed, reset, or replaced if an adversary gains sufficient
access. A living fungal seal, by contrast, cannot be reset to its original physiological
state. Physical disturbance permanently alters hyphal structure, growth patterns, and
electrical properties, creating persistent evidence of interference. Even if external
electronics are destroyed, the biological record remains embedded in the material itself.

\subsection{Energy requirements and long-duration operation}

Digital sensors and AI-enabled monitoring systems typically require continuous or
periodically replenished energy supplies. Batteries degrade, solar panels are vulnerable
to environmental conditions, and wired power may be unavailable or unreliable in remote
or disaster-prone environments. Energy constraints often impose trade-offs between sampling
rate, communication frequency, and system lifetime.

Fungal systems operate on radically different energy assumptions. Metabolic activity is
sustained by low-grade organic substrates and environmental nutrients, and electrical
signalling occurs at millivolt scales. While fungal systems are slow by digital standards,
they can operate continuously over months or years without external power infrastructure.
Energy-intensive operations such as data transmission and analysis can be deferred until
biologically significant changes occur.

A practical example is long-term environmental monitoring in remote locations. Deploying
and maintaining electronic sensor networks in deserts, polar regions, or dense forests is
logistically complex and costly. A fungal substrate embedded in soil or structural material
can persist with minimal intervention, adapting to seasonal variation while retaining
sensitivity to anomalous events such as contamination, excavation, or mechanical stress.
Periodic, low-bandwidth interrogation of the fungal system can replace continuous data
streams, reducing energy and communication demands.

\subsection{Temporal characteristics and information processing}

Digital systems excel at high-speed sampling, precise timing, and rapid response. These
capabilities are essential for control systems, real-time analytics, and safety interlocks.
However, they also encourage architectures optimised for immediacy not persistence.
Short-term events dominate attention, while slow-onset processes may be overlooked or
filtered out.

Fungal systems operate on slower timescales, integrating information over minutes, hours,
or days. Electrical spikes, oscillatory patterns, and morphological changes encode temporal
structure in a distributed and analogue manner. This makes fungal substrates particularly
well suited to detecting gradual processes such as fatigue, creep, subsidence, or chronic
environmental change.

As an example, consider subsurface intrusion detection. Digital systems often rely on
seismic sensors or active interrogation methods that may generate false alarms or require
frequent recalibration. A fungal network embedded in soil can integrate weak mechanical
disturbances over extended periods, producing cumulative physiological changes that signal
persistent activity. Such detection is inherently conservative,
prioritising long-term significance over immediacy.

\subsection{Security, attack surfaces, and cyber risk}

Digital monitoring systems introduce cyber-physical attack surfaces~\cite{loukas2015cyber,ashok2017cyber,duo2022survey}. Networked sensors,
edge devices, and AI pipelines can be compromised through software vulnerabilities,
malicious updates, spoofed data, or denial-of-service attacks~\cite{alotaibi2023survey,humayun2024securing,schmitt2023securing}. Securing these systems
requires continuous patching, authentication mechanisms, and active monitoring.

Fungal systems dramatically reduce cyber attack surfaces. The biological substrate itself
is not programmable in the conventional sense and does not execute code. While interfaces
between fungi and electronics must be secured, the core sensing and adaptive functions are
embodied in physical and physiological processes that cannot be hacked remotely. This
creates a form of security-through-orthogonality.

In high-risk facilities where cyber compromise is a primary concern, fungal sensing layers
can provide independent corroboration of physical events. Even if digital systems are
spoofed or disabled, biologically encoded evidence of disturbance or damage persists,
supporting forensic analysis and recovery.

\subsection{Integration and complementarity}

The comparison above highlights that fungal and conventional technologies excel under
different assumptions. Digital systems provide speed, precision, scalability, and rich
data processing, but are brittle under sustained disruption. Fungal systems provide
persistence, adaptability, and low observability, but operate slowly and produce signals
that require interpretation.

The most effective architectures therefore combine both approaches. Fungal substrates can
serve as adaptive front-end filters, embedding sensing, baseline learning, and tamper
evidence directly into materials and environments. Conventional electronics can perform
intermittent readout, logging, interpretation, and integration with human decision-making
processes. Such hybrid systems reduce reliance on continuous connectivity and computation
while preserving the strengths of digital analysis when it is available.

Instead of asking whether fungal systems can outperform conventional technologies, the
more productive question is how they can fail differently. In resilience engineering,
systems that fail differently --- and independently --- are more valuable than systems that fail
better under ideal conditions. From this perspective, fungal systems offer a unique and
complementary pathway toward security and resilience in environments where disruption is
not an exception, but the norm.

\section{Discussion and conclusion}

We argued that living fungal systems offer a complimentary approach to security and resilience than conventional digital technologies. Instead of prioritising speed, precision, and centralised control, fungal systems embody principles of persistence, adaptation, decentralisation, and
graceful degradation. These principles are not imposed through design but emerge naturally from fungal biology, making mycelial networks particularly well suited to operation in disrupted, uncertain, and physically hostile environments.

Throughout the paper, we have shown how fungal properties—distributed architecture, electrophysiological signalling, embodied memory, self-healing, and low observability—map directly onto the requirements of resilient security systems. Living mycelium can function simultaneously as sensing substrate,
protective material, and adaptive filter, blurring the  conventional separation between structure, sensor, and processor. Importantly, fungal systems do not aim to replace existing digital technologies. Instead, they provide an orthogonal
layer of awareness and protection that continues to function when conventional systems degrade, fail, or are deliberately disrupted.

From a practical perspective, the development of fungal systems for security and resilience requires a staged and cautious research trajectory. Initial efforts must focus on controlled laboratory characterisation, mapping electrophysiological
responses to well-defined mechanical, chemical, optical, and electrical stimuli. Key challenges include quantifying signal stability, understanding long-term drift, and establishing minimal metrics for sensitivity, selectivity, and persistence. Such characterisation is essential to distinguish biologically
meaningful signals from environmental noise and to define operational envelopes for deployment.

Subsequent work should emphasise biohybrid prototyping, where fungal substrates are interfaced with minimal electronic systems for signal acquisition and interpretation. These prototypes should prioritise robustness and repeatability over performance, using simple thresholding or pattern-based approaches rather
than complex machine learning pipelines. Controlled growth environments and standardised interfaces will be necessary to manage biological variability while preserving adaptive capacity.

Ultimately, the value of fungal systems must be assessed through controlled field trials embedded in representative materials and environments. Long-duration tests are particularly important, as fungal advantages emerge over time not in short-term demonstrations. Field trials should evaluate not only sensing
performance but also maintenance requirements, survivability under damage, and the persistence of tamper evidence under realistic stress scenarios.

At a conceptual level, fungal systems challenge prevailing assumptions about where sensing, memory, and decision support must reside. By embedding these functions directly into living materials, they enable a shift from reactive, event-driven monitoring toward persistent, adaptive awareness. This shift is
especially relevant in contexts where continuous connectivity, power, or human oversight cannot be guaranteed. Fungal systems function not as alarms or controllers, but as witnesses—maintaining embodied records of environmental conditions, stress, and interference over extended periods.

What are limitations, risks, and practical considerations of using living mycelium for security and resilience? 

While fungal systems offer distinctive advantages for security and resilience, their limitations and risks must be acknowledged. A primary constraint is growth speed. Mycelial networks operate on timescales of hours to days, making them unsuitable for applications requiring immediate response or fine-grained temporal resolution. Their strengths instead lie in persistent monitoring, long-term adaptation, and integration of slow-onset processes.

Environmental dependency represents a further limitation. Fungal viability and signalling are influenced by temperature, moisture, substrate composition, and nutrient availability. Extreme or rapidly fluctuating conditions may suppress activity or alter baseline behaviour. Practical deployments therefore require careful selection of species, substrates, and encapsulation strategies to ensure stability within defined operational envelopes.

Ethical and biosafety considerations must also be addressed. The use of living systems in security and infrastructure contexts raises questions regarding containment, unintended ecological interaction, and long-term environmental impact. These risks can be mitigated through the use of non-pathogenic species, physical containment within composite materials, growth limitation strategies, and compliance with existing biosafety and environmental regulations. Importantly, the proposed systems do not rely on genetic modification or uncontrolled release, reducing regulatory and ethical complexity.

Acknowledging these limitations does not diminish the value of fungal systems for resilience-oriented applications. Rather, it clarifies the contexts in which such systems are appropriate and highlights the need for cautious, staged development grounded in laboratory characterisation, controlled field trials, and transparent oversight.

By integrating sensing, adaptation, and repair within living substrates, fungal systems offer a complementary resilience paradigm—one grounded in endurance, recovery, and embodied intelligence. Future work should prioritise rigorous validation, careful hybridisation with digital systems, and transparent ethical oversight. If successful, fungal systems may enable protective infrastructures that do not merely resist disruption, but persist through it.

\section*{Acknowledgements}

The research has been conducted under the framework of the FUNGATERIA (\url{www.fungateria.eu}) project, which has received funding from the European Union’s HORIZON-EIC-2021-PATHFINDER CHALLENGES programme under grant agreement No. 101071145. It is co-funded by the UK Research and Innovation grant No. 10048406.

\end{document}